\documentclass[prl,twocolumn,superscriptaddress,noshowpacs,notitlepage,longbibliography]{revtex4-1}%
\usepackage{graphicx,bm,times}
\usepackage{amsmath}
\usepackage{amsfonts}
\usepackage{amssymb}
\usepackage{color}
\usepackage{hyperref}
\hypersetup{
	colorlinks = true,
	allcolors = {blue}
}

\begin{document}
	
	\title{$k_z$ selective scattering within Quasiparticle Interference measurements of FeSe}
	
	\author{Luke C. Rhodes}
	\affiliation{Department of Physics, Royal Holloway, University of London, Egham, Surrey, TW20 0EX, United Kingdom}
	\affiliation{Diamond Light Source, Harwell Campus, Didcot, OX11 0DE, United Kingdom}
	\affiliation{School of Physics and Astronomy, University of St. Andrews, St. Andrews KY16 9SS, United Kingdom}	
	
	\author{Matthew D. Watson}
	\affiliation{School of Physics and Astronomy, University of St. Andrews, St. Andrews KY16 9SS, United Kingdom}	
	
	\author{Timur K. Kim}
	\affiliation{Diamond Light Source, Harwell Campus, Didcot, OX11 0DE, United Kingdom}

	\author{Matthias Eschrig}
	\affiliation{Department of Physics, Royal Holloway, University of London, Egham, Surrey, TW20 0EX, United Kingdom}
	\affiliation{Institute of Physics, University of Greifswald, Felix-Hausdorff-Strasse 6, 17489 Greifswald, Germany}

	\begin{abstract}
	Quasiparticle interference (QPI) provides a wealth of information relating to the electronic structure of a material. However, it is often assumed that this information is constrained to two-dimensional electronic states.
	Here, we show that this is not necessarily the case. For FeSe, a system dominated by surface defects, we show that it is actually all electronic states with negligible group velocity in the $z$ axis that are contained within the experimental data. By using a three-dimensional tight binding model of FeSe, fit to photoemission measurements, we directly reproduce the experimental QPI scattering dispersion, within a T-matrix formalism, by including both $k_z = 0$ and $k_z = \pi$ electronic states. This result unifies both tunnelling and photoemission based experiments on FeSe and highlights the importance of $k_z$ within surface sensitive measurements of QPI. 
	\end{abstract}
	\date{\today}
	\maketitle
	
	
The  iron-based superconductor FeSe has recently been a focal point in the study of unconventional superconductivity. The momentum dependence of the superconducting gap, extracted from angle-resolved photoemission spectroscopy (ARPES) \cite{Xu2016,Hashimoto2018,Liu2018,Kushnirenko2018,Rhodes2018} and QPI measurements \cite{Sprau2017,Hanaguri2018}, has been shown to be highly two-fold symmetric, and sensitive to the orbital content of the bands. However, a consensus relating to the full theoretical implications of this gap structure have remained limited due to the range of different, and often contradictory, models of the electronic structure used as a starting point for theoretical investigations \cite{Sprau2017,Kreisel2017,Rhodes2018,Benfatto2018,Kang2018,Hu2018}. 

In order to resolve the differences in theoretical models of the electronic structure, it is important to study the results and conclusions extracted from experimental measurements, such as ARPES and QPI. However, there is currently a discrepancy between the interpretation of the data obtained by these two techniques. QPI measurements of FeSe \cite{Kasahara2014,Kostin2018,Hanaguri2018}, obtained via scanning tunnelling microscopy (STM), have been interpreted as consistent with a theoretical model where the Fermi surface consists of one hole pocket, two electron pockets, and exhibits a large difference in the quasiparticle weight of the $d_{xz}$ and $d_{yz}$ orbitals \cite{Kostin2018}. Whereas, the orbital sensitive measurements from ARPES have been interpreted as consistent with a model with roughly equivalent quasiparticle weights for the $d_{xz}$ and $d_{yz}$ orbitals \cite{Fanfarillo2016,Liu2018,Fedorov2016,Rhodes2018}, but with only one hole pocket and one electron pocket at the Fermi surface \cite{Watson2017c,Yi2019_arXiv,Huh2019_arXiv}. 
Interestingly, both of these interpretations have been used to correctly describe the momentum dependence of the superconducting gap \cite{Kreisel2017,Rhodes2018}, yet completely contradict one another on the topology and orbital coherence of the electronic structure. 

In this report, we address this discrepancy. We show that the ARPES-based interpretation of the electronic structure, i.e a model which has equal quasiparticle weights for all orbitals, but only has one hole pocket and one electron pocket at the Fermi level, is fully consistent with the QPI measurements of FeSe, once all electronic scattering vectors which exhibit zero Fermi velocity in the $k_z$ axis are taken into account. This result therefore provides experimental unification of the electronic structure of FeSe, as determined by ARPES and STM, and highlights the importance of $k_z$ within the surface sensitive measurements of QPI.
 
 	 \begin{figure}
 	 	\centering
 	 	\includegraphics[width = 0.97\linewidth]{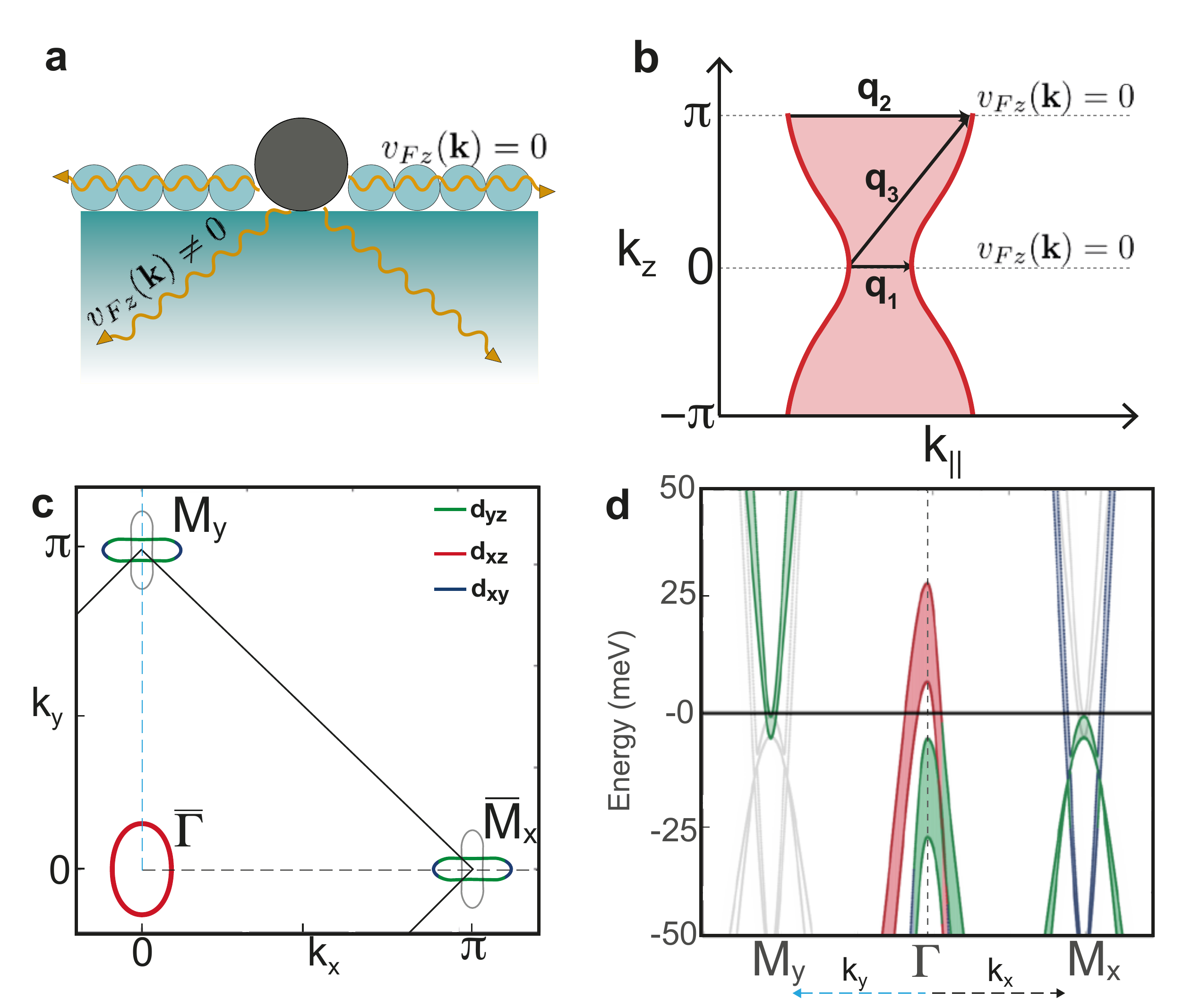}
 	 	\caption{a) Electron propagation from a surface defect, STM is only sensitive to the top-most layer of the material, therefore, only QPI arising from electronic states with $v_{Fz}(\mathbf{k}) = 0$ (orange arrows) will be detected via STM. b) Cut of the $k_z$ dispersion of the hole pocket of FeSe, highlighting states where $v_{Fz}(\mathbf{k}) = 0$. This then leads to three sets of scattering vectors with unique $k_z$ dependence, labelled as $\mathbf{q_1,q_2,q_3}$, This description equivalently holds for the electron pocket and interpocket scattering. c) Fermi surface of the ARPES-based model of FeSe at $k_z = \pi$. The black line describes the Brillouin zone boundary. d) Band dispersion along $\bar{M}_y-\bar{\Gamma}-\bar{M}_x$ with projected $k_z$ states. The solid lines describe $k_z = 0$ and $k_z =\pi$ states, whereas the shaded region indicate states with intermediate $k_z$. The grey bands show the states that we have manually excluded from the calculation in order to reproduce ARPES measurements \cite{Watson2017c}.}
 	 	\label{fig:fig1}
 	 \end{figure}

\section{Methodology}
QPI is the phenomenon which occurs when propagating quasiparticles scatter elastically from a local potential caused by an impurity, leading to interference patterns in the local density of states. This effect is observable in an STM measurement as oscillations in the spatial-dependence of the tunnelling current emanating from a defect or vacancy. The Fourier transform of these oscillations then unveils a momentum- and energy-dependent structure that may be interpreted in terms of scattering processes, $\mathbf{q} = \mathbf{k}-\mathbf{k'}$, where the momentum $\mathbf{k}$ and $\mathbf{k'}$ are defined by the underlying single particle electronic structure. However, it is important to note that the information obtained from this two dimensional Fourier transform comes from measurements which, in real space, are restricted to tunnelling from the top-most layer of the material. As a consequence, for defects located in the surface layer of a material, which is the dominant form of impurities in FeSe \cite{Sprau2017}, this imposes a restriction: Any coherent oscillations detected far away from a defect must arise from electronic states which have a vanishing group velocity in the $z$ axis. We illustrate this principle in Fig. \ref{fig:fig1}(a).

This condition restricts the detectable scattering processes, to electronic states where both $\mathbf{k}$ and $\mathbf{k'}$ have zero Fermi velocity along the $k_z$ axis \cite{Weismann2009,Lounis2011}, which, in the case of FeSe is true for both $k_z = 0$ and $k_z = \pi$ states. Thus, there should be three sets of scattering vectors, labelled as $\mathbf{q_1,q_2}$ and $\mathbf{q_3}$ in Fig. \ref{fig:fig1}(b), which will produce detectable QPI within STM measurements. Previous QPI calculations of FeSe have included the $k_z = 0$ to $k_z = 0$ set of scattering vectors ($\mathbf{q_1}$) \cite{Kostin2018,Singh2018} but have neglected the contributions from $\mathbf{q_2}$ and $\mathbf{q_3}$. In this work, we include all three sets of $\mathbf{q}$ vectors within the typical T-matrix calculation of the local density of state (LDOS, $N(\mathbf{q},\omega)$) \footnote{For details, please refer to the Supplemental Material}, where the summation over both $k_z$ and $q_z$ are constrained to the 0 or $\pi$ plane and the total LDOS is then the sum of the three sets of $\mathbf{q}$ vectors,

\begin{equation}
\label{Eq:LDOS}
\tilde{N}(q_x,q_y,\omega) = \sum_{q_z\in [0,\pi]} N(q_x,q_y,q_z,\omega).
\end{equation}

We note that Eq. \eqref{Eq:LDOS} is calculated assuming bulk QPI scattering. Formally, open boundary conditions in the $z$ axis should be used to describe the surface of a material \cite{Zhou2019}, however, as no surface states have been detected for FeSe \cite{Coldea2018}, Eq. \eqref{Eq:LDOS} is still a valid approximation, which, as we will show, accurately describes the experimental data.

In order to facilitate a comparison between theory and experimental data, we next calculate the normalised LDOS, $L(\mathbf{r},\omega)$, also referred to as the Feenstra function \cite{Feenstra1994},

\begin{equation}
\label{Feenstra_main}
L(\mathbf{r},\omega) = \frac{\tilde{N}(\mathbf{r},\omega)}{\sum_{\omega'=0}^{\omega'=\omega}\tilde{N}(\mathbf{r},\omega')}.
\end{equation}

Here, $\tilde{N}(\mathbf{r},\omega)$ is the 2D inverse Fourier transform of Eq. \eqref{Eq:LDOS}. We then plot the Fourier transform of $L(\mathbf{r},\omega)$, in moment space, and directly compare our results with the QPI data from Ref. \cite{Hanaguri2018}.

To calculate Eq. \eqref{Feenstra_main}, we employ a tight binding model which has been optimised to describe the band dispersions determined from ARPES measurements of detwinned crystals of FeSe \cite{Watson2017c}. In Ref. \cite{Watson2017c}, it was observed that, at low temperatures, the Fermi surface of FeSe consisted of one hole pocket and a single electron pocket. However, tight binding models of FeSe suggest that two electron pockets should be present at the Fermi surface \cite{Mukherjee2015}. To account for this experimental observation, we have chosen to specifically exclude bands associated with this unobserved second electron pocket. These bands are shown in grey in Fig. \ref{fig:fig1}(c,d). To exclude these bands, we use the unfolded one-Fe unit cell of FeSe, which separates the electron pockets in momentum space \cite{Brouet2012,Nica2015}. We then employ a Green's function,

\begin{equation}
\hat{G_0}(\mathbf{k},\omega) = \frac{1}{(\omega + i\Gamma(\mathbf{k}))\hat{1}-\hat{H^0}(\mathbf{k})},
\label{Eq:Green's Function}
\end{equation}

\noindent that includes a momentum dependent broadening parameter, $\Gamma(\mathbf{k})$. We define $\Gamma(\mathbf{k})$ as arbitrarily large ($>100$~eV) in the vicinity of $\mathbf{k} =(0,\pi)$, where our tight binding model incorrectly describes the presence of a second electron pocket. For all other momenta we set $\Gamma(\mathbf{k})$ to 1.25~meV. The Hamiltonian, $\hat{H^0}(\mathbf{k})$, then describes hopping between all five d-orbitals in the presence of spin orbit coupling. This is discussed in detail in Ref. \cite{Rhodes2017,Rhodes2018}. In this report we have additionally reduced the contribution of spin orbit coupling at the $M$ point to improve the agreement of the band positions at negative energies. To do this, we have reduced the spin orbit coupling strength of the $l_x$ and $l_y$ components to $\lambda_{x/y} = 5$~meV. $\lambda_z$ is then set to 19~meV. This form of the spin orbit coupling matrix is defined in \cite{Saito2015a}. The removal of states ``by hand", makes this model highly phenomenological, however, within the energy region of $\pm50$~meV, this approach quantitatively reproduces the experimental band dispersions measured by ARPES experiments on detwinned crystals \cite{Watson2017c,Yi2019_arXiv,Huh2019_arXiv}. 
 
\begin{figure*}
	\centering
	\includegraphics[width=\linewidth]{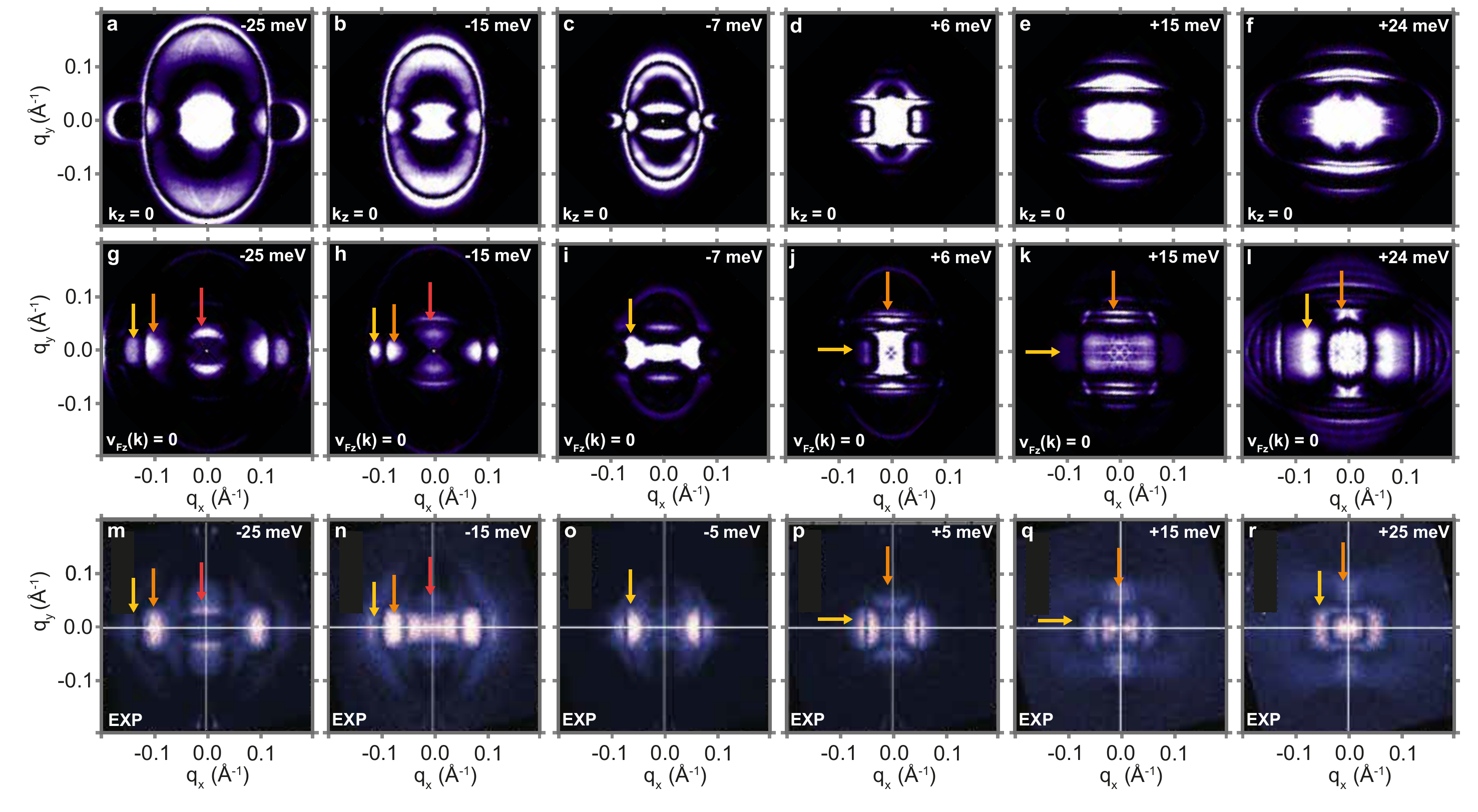}
	\caption{ a-f) Calculated Feenstra function, $|L(\mathbf{q},\omega)|$ for a two-dimensional, $k_z = 0$ model of FeSe at various energies. g-i) Equivalent calculations for a model of FeSe which includes both $k_z = 0$ and $k_z = \pi$ states ($v_{Fz}(\mathbf{k}) = 0$). m-r) Experimental QPI data. Adapted from Ref. \cite{Hanaguri2018} under the Creative Commons Attribution 4.0 International License. The arrows highlight features which are observed in experiment and captured within this theoretical framework.}
	\label{fig:fig2}
\end{figure*}

\begin{figure}
	\centering
	\includegraphics[width=\linewidth]{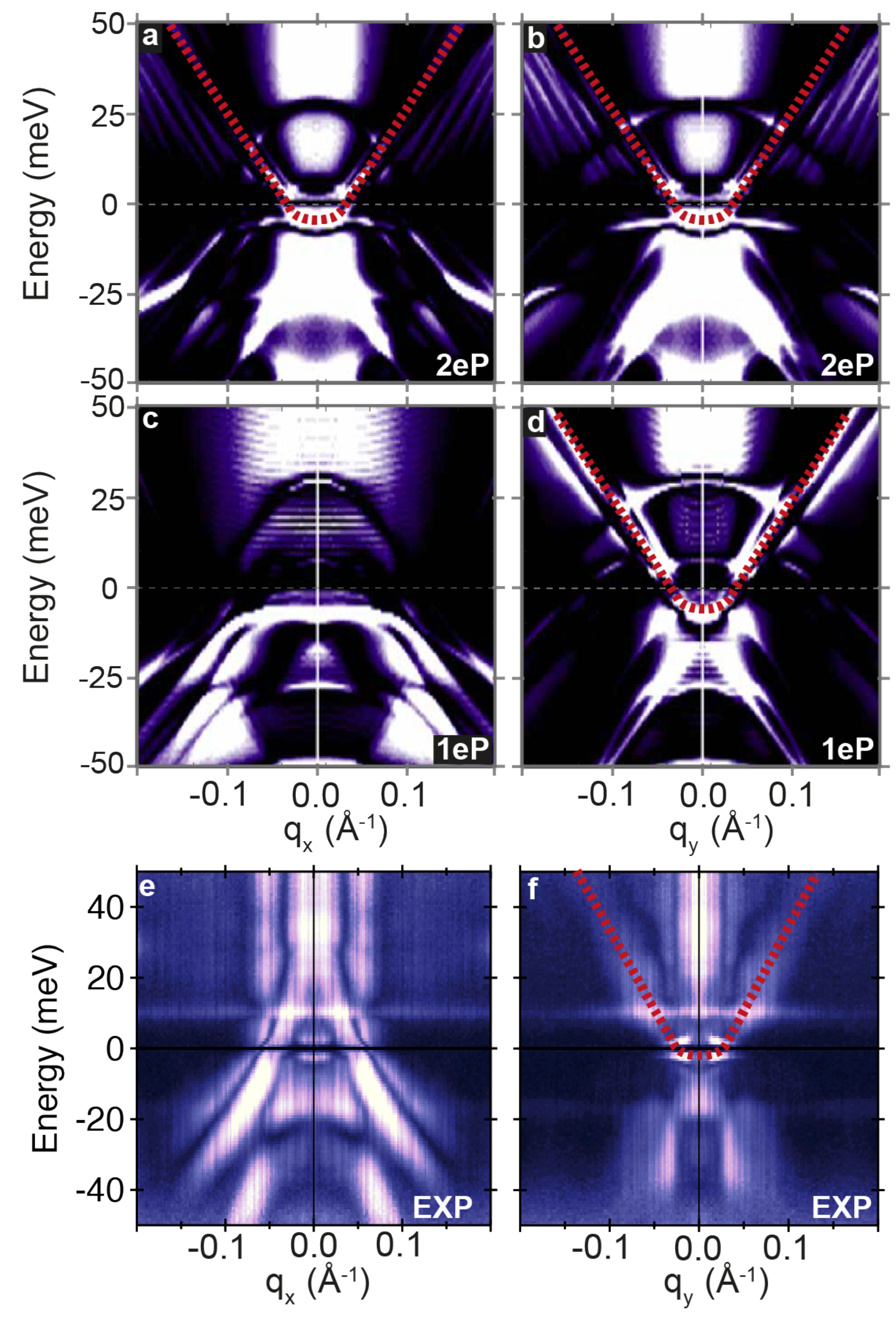}
	\caption{Energy vs momentum QPI based band dispersions. a,b) QPI band dispersions along $q_x$ at $q_y = 0$ and $q_y$ at $q_x$ = 0 for a ``two-electron-pocket" (2eP) model of FeSe. The dashed red line highlights electron like dispersions. c,d) Equivalent QPI band dispersion for a ``one-electron-pocket" (1eP) model. e,f) Experimental QPI band dispersions. Adapted from Ref. \cite{Hanaguri2018} under the Creative Commons Attribution 4.0 International License. }
	\label{fig:fig3}
\end{figure}

\section{Results}
In Fig. \ref{fig:fig2}, we show the impact of including all electronic states with $v_{Fz}(\mathbf{k}) = 0$ within the modelled QPI dispersions. In Fig. \ref{fig:fig2}(a-c) we present the calculated result for a $k_z = 0$ model of FeSe at negative energies. Here, the scattering vectors are dominated by the elliptical outer hole band, which decreases in radius as the energy approaches the Fermi level. When all states with $v_F(z) = 0$ are included, however, shown in Fig. \ref{fig:fig2}(g-i), the intensity of the scattering vectors, associated with the $k_z  = 0$ hole band, is suppressed. Conversely, the intensity from the scattering vectors associated with the two-dimensional bands around the $M$ point is enhanced. This results in a highly anisotropic scattering dispersion, parallel to the $q_x$ axis, which is in very good agreement with the experimental results of Ref. \cite{Hanaguri2018}, shown in Fig. \ref{fig:fig2}(m-o). In fact, all the scattering vectors observed at these energies are accounted for, with the exception of the central scattering vector in Fig. \ref{fig:fig2}(n), which is likely a limitation of our tight binding parameterisation.

Direct agreement between theory and experiment can also be found at positive energies. The $k_z = 0$ model of FeSe exhibits a hole band maxima at +7~meV \cite{Watson2017,Liu2018,Hashimoto2018}. This means that for energies greater than +7~meV, only a single elliptical electron band will contribute to the scattering dispersion in our model. This is shown in Fig. \ref{fig:fig2}(d-f). Scattering vectors arising from the $k_z = 0$ hole band can be observed in Fig. \ref{fig:fig2}(d), at $\omega = +6$~meV, however, for $\omega = +15$ and +24~meV only an elliptical dispersion from the electron band is observed. Alternatively, at $k_z = \pi$ the hole band captured within the model has a maximum value of +27~meV. Thus, the inclusion of $k_z = \pi$ electronic states, (via the inclusion of $\mathbf{q_2}$ and $\mathbf{q_3}$ from Fig. \ref{fig:fig1}b) adds scattering dispersions associated with both hole and electron states. This produces the scattering dispersion shown in Fig. \ref{fig:fig2}(j-l) which is in much better agreement with the experimental measurements of Ref. \cite{Hanaguri2018}, shown in Fig. \ref{fig:fig2}(p-r). In Fig. \ref{fig:fig2}(j), it is noted that the intensity of the scattering vectors appears rotated, compared to the experimental measurement of Ref. \cite{Hanaguri2018}, shown in Fig. \ref{fig:fig2}(p). This intensity difference can arise from anisotropic scattering processes, which are not included in this calculation. However, as the length of the scattering vectors are correctly described, we conclude that the QPI measurements of FeSe are sensitive to all electronic states with $v_{Fz}(\mathbf{k}) = 0$.

In Fig \ref{fig:fig2}, we have obtained very good agreement with the experimental QPI measurements of Ref. \cite{Hanaguri2018} on the assumption that the Fermi surface of FeSe consists of one hole pocket and one electron pocket, as determined by ARPES studies on detwinned crystals \cite{Watson2017c,Yi2019_arXiv,Huh2019_arXiv}. However, as discussed previously, ab-initio calculations \cite{Eschrig2009,Fedorov2016}, and most theoretical models of FeSe, suggest that the Fermi surface should consist of one hole pocket and two electron pockets. To further support the ``one-electron pocket" scenario of FeSe, we now focus on the band dispersions determined from QPI. If we include both electron-pockets into the calculation, by including the bands shown in grey in Fig. \ref{fig:fig1}(c), we find that the QPI derived band dispersions along the $q_x$ and $q_y$ axis are predicted to be very similar, as shown in Fig. \ref{fig:fig3}(a,b). In particular, the ``two-electron-pocket" scenario predicts electron-like dispersions in both the $q_x$ and $q_y$ directions. This is in stark contrast to the experimental measurements of Ref. \cite{Hanaguri2018}, shown in Fig. \ref{fig:fig3}(e-f), where electron-like dispersions are only observed along the $q_y$ axis. When we repeat this calculation using the ``one-electron-pocket" model of FeSe, shown in Fig. \ref{fig:fig3}(c,d), we indeed correctly reproduce this anisotropic scattering dispersion, with electron-like dispersions only present along the $q_y$ axis. Moreover, below the Fermi level, in the ``one-electron-pocket" model, more hole like bands are predicted to disperse along the $q_x$ axis than the $q_y$ axis, which is exactly what is observed in the experimental measurements of Fig. \ref{fig:fig3}(e,f). From this, we conclude that QPI measurements are in agreement with the electronic structure determined by ARPES measurements, where only one electron pocket is detected at the Fermi surface.

\section{Discussion}
In this manuscript we have shown that, in a system dominated by surface defects, it is not only the $k_z =0$ electronic states that are detected by QPI measurements, but all states with zero Fermi velocity in $k_z$. We have focused on the case of FeSe, however, the conclusions drawn about the nature of scattering should be general for many materials, including heavy Fermions \cite{Zhou2013} and other pnictides \cite{Chi2014}. 

There is great current interest in the fate of the second electron pocket of FeSe, which is predicted to exist within the nematic phase \cite{Mukherjee2015,Jiang2016} but is mysteriously not observed by ARPES at low temperatures \cite{Watson2017c,Yi2019_arXiv,Huh2019_arXiv}. Recently, it has been proposed that this missing electron band is actually pushed above the Fermi level, driven by a particular hybridisation scheme at the zone boundary in the nematic phase \cite{Yi2019_arXiv,Huh2019_arXiv}. However, in our understanding, the QPI data does not support this interpretation; there is no band minimum observed above the Fermi level, and no low-$q$ electron-like dispersions in the $q_x$ direction (Fig 3e).
Another proposition is that the apparent absence of spectral weight on the second electron pocket is a manifestation of orbital-selective quasiparticle weights \cite{Kostin2018}, since the phenomenological suppression of $d_{xz}$ and $d_{xy}$ weight would particularly affect this pocket. In the supplementary material, we present simulations including the quasiparticle weight factors suggested in Refs. \cite{Kostin2018}. We find that, although this approach does account for the general observation of highly two-fold symmetric QPI dispersion and qualitatively captures some features, a more satisfactory agreement can be found within our approach, in which we keep all the orbitals coherent but implement a pocket-selective coherence.

In this work, our approach has been to take the ARPES data at face value, and thus to phenomenologically exclude the second electron pocket, and its associated bands, despite the fact that they are present in any reasonable tight-binding model of FeSe. Under this assumption, our simulations correctly reproduce many features of the QPI data, which is not the case when the second electron pocket is included. Thus, this technique, which is independent and complementary to that of ARPES, seems to also indicate the presence of only one electron pocket at the Fermi surface of FeSe.  Previously, we have argued that the ``one-electron-pocket" scenario can also naturally account for the observed superconducting gap structure of FeSe \cite{Rhodes2018}, which has been further supported by specific heat measurements \cite{Hardy2018}.  There is, therefore, mounting experimental support for the ``one-electron pocket" scenario of FeSe, which calls for further theoretical and experimental investigations to elucidate the origin of this effect.

\begin{acknowledgments}
	We thank T. Hanaguri for his insight and for providing high quality images of previously published experimental data \cite{Hanaguri2018} for comparison with this work. We would also like to thank A. Chubukov, A. I. Coldea, A. A. Haghighirad, P. D. C. King, M. van Schilfgaarde, C. Trainer, P. Wahl, M. Yi and C. M. Yim for useful discussions. 
\end{acknowledgments}

\clearpage

\onecolumngrid
\appendix

\section{Theoretical methodology}
To model the QPI scattering dispersions, we calculate the density of states in the presence of a non-magnetic impurity,

\begin{equation}
	\label{LDOS}
	N(\mathbf{r},\omega) = N_0(\omega) + \delta N(\mathbf{r},\omega).
\end{equation}

Here, $N_0(\omega)$ is the unperturbed density of states,
\begin{equation}
	N_0(\omega) = -\frac{1}{\pi}Tr \big[Im \sum_{k_x,k_y}\sum_{k_z \in[0,\pi]} \hat{G_0}(\mathbf{\mathbf{k}},\omega)\big].
\end{equation}

Here, $\mathbf{k} = \{k_x,k_y,k_z\}$ is a 3D vector with and $\hat{G_0}$ is the Green's function described in Eq. 3 of the main text.

To calculate the local density of states (LDOS), we assume a single non-magnetic impurity that can be described by a point-like delta function. Under this approximation, the perturbation to the non-interacting density of states, for a centrosymmetric system, may be calculated in momentum space as,

\begin{equation}
	\delta N(\mathbf{q},\omega) = -\frac{1}{\pi}Tr\Big[ Im \sum_{k_x,k_y}\sum_{k_z \in[0,\pi]} \hat{G_0}(\mathbf{k},\omega)\hat{T}(\omega)\hat{G_0}(\mathbf{k+q},\omega) \Big].
	\label{Eq:dNq}
\end{equation}
Here, we restrict the integral in $k_z$ to the values which have a stationary curvature, and thus zero Fermi velocity in the $k_z$ axis, as discussed in the main text. Eq. \eqref{Eq:dNq} can then be inverse Fourier transformed to obtain $\delta N(\mathbf{r},\omega)$. Here, $\hat{T}(\omega)$ is the T-matrix which describes all scattering processes associated with the impurity,

\begin{equation}
	\hat{T}(\omega) = \frac{V}{ \hat{1}-V\sum_{\mathbf{k}}\hat{G_0}(\mathbf{k},\omega)}.
\end{equation}

As we are assuming a point-like impurity, the impurity potential, $V$, is assumed to be a constant. In this report we set $V=-100$~meV, as suggested by tunnelling experiments on FeSe \cite{Kostin2018}.

\section{Comparison with orbital-selective quasiparticle weights}
Here we present a comparison of the calculated Feenstra function using the ''one-electron pocket" model, to a ''two-electron-pocket" model with the inclusion of orbital-selective quasiparticle weights proposed in Ref. \cite{Sprau2017,Kostin2018}. Following the methodology of Kostin et. al. \cite{Kostin2018}, we modify the Greens function to include anisotropic quasiparticle weights via,

\begin{equation}
	\tilde{G}^{nm}_{0}(\mathbf{k},\omega) = \sqrt{Z_n}\sqrt{Z_m}\hat{G}^{nm}_{0}(\mathbf{k},\omega) 
\end{equation}

\begin{figure*}
	\centering
	\includegraphics[width=0.95\linewidth]{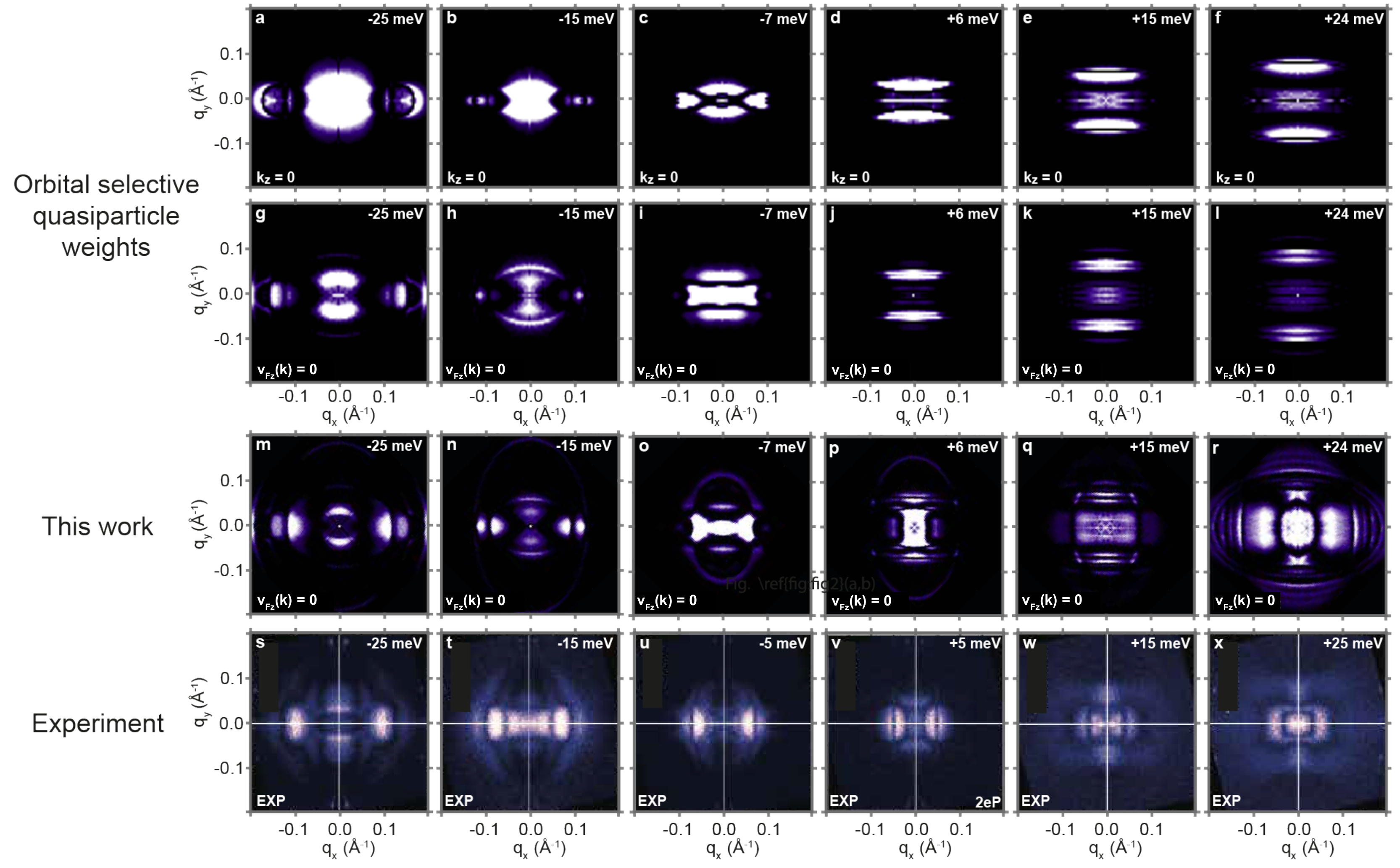}
	
	\caption{Comparison of the orbital-selective quasiparticle weight scenario with the ``one-electron pocket" scenario studied in this report. a-f) Calculation of $L(\mathbf{q},\omega)$ for the same energies as in the main text, assuming anisotropic Quasiparticle weights and only including $k_z = 0$ states, equivalent to the assumptions of Ref. \cite{Kostin2018}. g-l) The same orbital-selective quasiparticle weight calculation including all states with $v_{Fz}(\mathbf{k}) = 0$. m-r) $L(\mathbf{q},\omega)$, assuming a one-electron-pocket scenario and $v_{Fz}(\mathbf{k}) = 0$. Presented in Fig. 2(g-i) of the main text. s-x) Experimental data at equivalent energies. Reproduced from Ref. \cite{Hanaguri2018} under the Creative Commons Attribution 4.0 International License.}
	\label{fig:SM1}
\end{figure*}

The quasiparticle weights, $Z_{i}$, are then defined for each orbital,$i$ as $[d_{xy} = 0.073,d_{x^2-y^2}=0.94,d_{xz}=0.16,d_{yz}=0.85,d_{z^2}=0.36]$ equivalent to that of Ref. \cite{Kostin2018}. In Fig. \ref{fig:SM1} we compare the two approaches. In Fig. \ref{fig:SM1}(a-f) we present the calculated Feenstra function assuming orbital-selective quasiparticle weights and that only $k_z = 0$ states contribute to scattering. It can be seen that a large anisotropy is induced, which resembles the experimental scattering vectors at negative energies. However, at positive energies only two parallel sets of scattering vectors are predicted to occur, which arise from the $d_{yz}$ electron pocket. This does not agree with the experimental data from Ref. \cite{Hanaguri2018} where a more complicated scattering dispersion is presented. It can also be seen that including $k_z = \pi$ states into this assumption (shown in Fig. \ref{fig:SM1}(g-l)) does not improve the agreement at positive energies. We attribute this to the fact that scattering associated with the $d_{xz}$ hole band at $k_z = \pi$ is strongly suppressed.

\end{document}